# Scattering-coded elastic meta-boundary

Tianxi Jiang[1+], Xinxin Liao[1+], Hao Huang[1], Zhi-Ke Peng[1,2], Qingbo He[1*]

1. State Key Laboratory of Mechanical System and Vibration, Shanghai Jiao Tong University, 200240 Shanghai, People's Republic of China
2. School of Mechanical Engineering, Ningxia University, 750021 Yinchuan, People's Republic of China

\* Corresponding author. Email: qbhe@sjtu.edu.cn
+ These authors contributed equally to this work.

Object localization through active elastic waves is a crucial technology, but generally requires a transducer array with complex hardware. Although computational sensing has been demonstrated to be able to overcome the shortcomings of transducer array by merging artificially designed structures into sensing process, coding spatial elastic waves for active object identification is still a knowledge gap. Here we propose a scattering-coded elastic meta-boundary composed of randomly distributed scatterers for computational identification of objects with a single transducer. The multiple scattering effect of the meta-boundary introduces complexity into scattered fields to achieve a highly uncorrelated scattering coding of elastic waves, thereby eliminating the ambiguity of the object location information. We demonstrate that the locations of objects can be uniquely identified by using the scattering coding of our designed meta-boundary, delivering a design of meta-boundary touchscreen for human-machine interaction. The proposed scattering-coded meta-boundary opens up avenues for artificially designed boundaries with the capability of information coding and identification, and may provide important applications in wave sensing, such as structural monitoring, underwater detection, indoor localization, and biomedical imaging.

**Introduction**

Locating objects by means of waves reflected back to the emitter by the objects is a crucial technology learning from creatures and developed by humans [1-3]. Compared with electromagnetic waves, elastic waves have less attenuation in water and solid mediums. Object localization through elastic waves can be widely used in fields such as orientation [4], imaging [5], obstacle avoidance [6], structural health monitoring [7], and human-machine interaction [8]. Typically, object localization is achieved by using a transducer array to emit short ultrasonic waves. The ultrasonic waves can be reflected by objects and then recorded by the same transducer array. The time delay between transmission of the waves and detection of the echoes contains the location information of objects. By using advanced algorithms to process the received signals, the location of objects can thus be identified.

The object localization methods using ultrasonic transducer array and array signal processing techniques (e.g., phased array imaging, synthetic aperture imaging, and total focusing imaging) have achieved successfully applications in medical imaging and industrial detection [9-11]. However, these methods require a large number of transducers, complex control circuits and data acquisition systems, resulting in high hardware costs and high-power consumption. The performance of localization and imaging is also greatly related with the placements of transducers, which limits the applications in narrow areas. Other methods, such as time reversal methods with chaos cavities [12], can reduce the number of transducers to a certain extent, but require active scanning of the entire fields.

Let us review the echolocation and sound source localization of creatures. In addition to the complex nervous system, it can be noticed that some creatures (e.g., bats, dolphins, and insects) have the structured vocal and auditory systems [1, 13-15], which can modulate the sound waves to distinguish the orientation of objects beyond the limitations of spatial sampling law. This characteristic inspires us to develop strategies to reduce the sensing requirement for object localization. Recently, a concept of computational sensing combined with artificially designed structures has been demonstrated [16-18]. Here, the structures merge the sensing process by replacing part of the digital signal processing with modulation of the structures, thereby giving structural features to the signals. The information can thus be reconstructed by solving the inverse problems via computational algorithms. As a consequence, the number of measurements required to recover the signal of interest can be drastically reduced, potentially reducing the dimension and complexity of the sensing system.

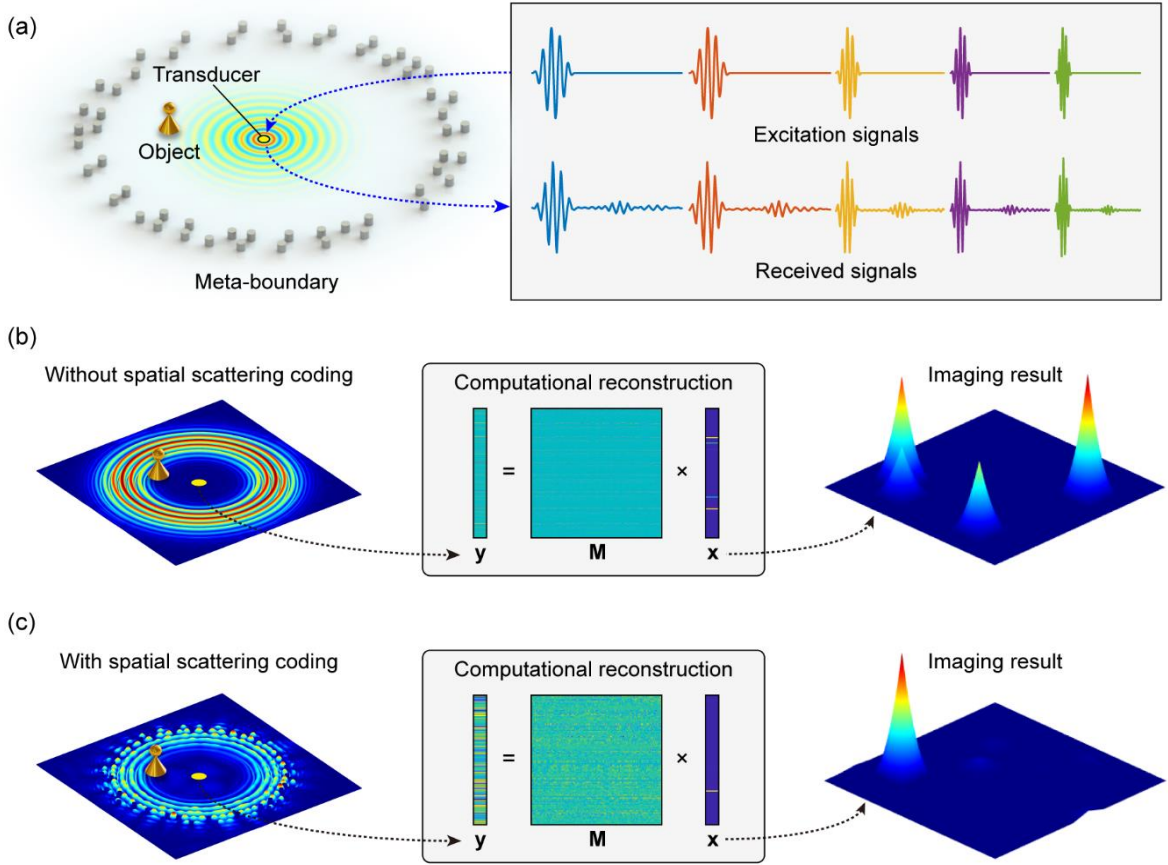

Figure 1. Illustration of scattering-coded elastic meta-boundary for computational identification with a single transducer. (a) Schematic diagram of the meta-boundary and the process of multiple measurements. (b) Schematic imaging result without spatial scattering coding. The location of the object may be incorrectly reconstructed at the symmetric points of the actual location. (c) By using the spatial scattering coding property of the meta-boundary, the object location can be uniquely identified.

One of the most impressive advances in computational sensing is the design of the single-pixel camera with compressive sensing [19]. This work has inspired many studies on computational imaging of electromagnetic and acoustic fields with only a single detector [20-22]. In these studies, the combination of compressive sensing and spatial coding structures, such as metamaterials, greatly reduces the requirement for sensing. Metamaterials are artificially designed structures with unusual effective physical properties that can demonstrate fascinating functionalities [23-29]. Specifically, metamaterials is capable of coding physical fields in temporal, spatial, and frequency domains, which is particularly suitable for computational sensing [30]. For instance, metamaterials with randomized structures can effectively break the uniform wave fields. Sub-wavelength acoustic imaging [31-33] and multiple source vibration identification [34, 35] have been achieved with only a single sensor by introducing the compressive sensing framework. These studies provide perspectives for passive source identification with cheaper, simpler, and smaller devices.

For active object localization, compared with the advances in electromagnetic waves [36, 37], computational identification with spatial coding metamaterials is rarely studied in elastic mediums, especially in solid mediums. Due to the symmetry of boundaries and the damping of materials, the object information carried by elastic waves is usually ambiguous and attenuated. How to code spatial elastic waves for active object identification is still a knowledge gap. In this work, we propose a concept of scattering-coded elastic meta-boundary for computational identification of elastic waves. The meta-boundary is composed of randomly distributed scatterers. We theoretically demonstrate that the meta-boundary can introduce complexity into the wave fields through multiple scattering to achieve highly uncorrelated scattering coding of elastic waves, thereby eliminating the ambiguity of object information. This property ensures that the locations of objects can be uniquely identified with only a single transducer by using the computational sensing framework. We further design a type of meta-boundary touchscreen for interactive input as a demonstration. Our study not only has potential application prospects in the fields of structural health monitoring and human-machine interaction, but also can provide exciting perspectives for underwater detection, indoor localization, and biomedical imaging.

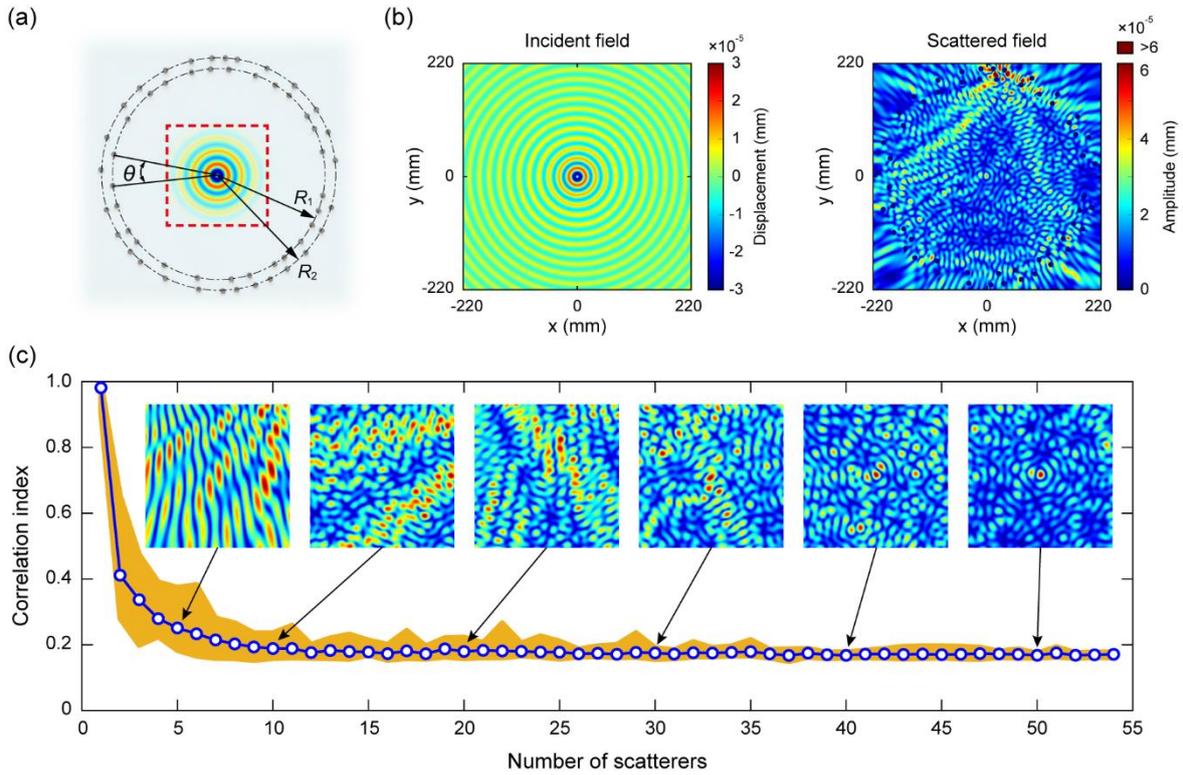

Figure 2. Theoretical analysis of spatial scattering coding with meta-boundary by using the multiple scattering theory. (a) Illustration of the meta-boundary in the theoretical analysis. The point source with the frequency of 40 kHz is excited at the center of the plate. (b) The incident and scattered fields calculated by using the multiple scattering theory. (c) The correlation index of the scattered fields with different number of scatterers. The dots represent the average of 10 calculations. The area around the dots represents the variation range of the correlation index in 10 calculations. The scattered fields corresponding to different number of scatterers are also illustrated.

**Results**

**Concept of the scattering-coded elastic meta-boundary.** The illustration of scattering-coded elastic meta-boundary for computational identification is shown in Fig. 1. The meta-boundary is composed of randomly distributed cylindrical scatterers clamped on a thin plate (Fig. 1(a)). Elastic waves are excited by a single transducer, propagate in the thin plate, and scatter when encountering an object. The scattered waves can be reflected by the meta-boundary and then picked up by the transducer. We regard the emission and reception of an excitation signal as one measurement of the object. The excitation signals include multiple pulses with different frequencies, which means that multiple measurements can be performed on the observation area in a wide frequency band for the subsequent computational identification. The computational identification process is expressed as $\mathbf{y} = \mathbf{M}\mathbf{x}$, where $\mathbf{y}$ is the measured signal from the single transducer, $\mathbf{M}$ is the measurement matrix, $\mathbf{x}$ is the objective vector containing the information of the objects. By using the reconstruction algorithm, the objective vector $\mathbf{x}$ can be solved to realize object identification.

The performance of identification mainly depends on the quality of the measurement matrix, which is determined by the spatial scattering coding characteristics of elastic waves. Figure 1(b) shows a schematic identification result of an object without spatial scattering coding. The location of the object is incorrectly reconstructed at the symmetric points of the actual location, because only a single transducer is incapable of distinguishing the ambiguous information carried by reflection waves. In contrast, the randomized meta-boundary can achieve the multiple scattering of elastic waves, breaking the spatial symmetry and introducing complexity into the wave fields (Fig. 1(c)). The spatial scattering coding of elastic waves can remove the ambiguity of the unmodulated scattered signals. Therefore, we can uniquely identify the spatial location of the object by using the computational reconstruction algorithm.

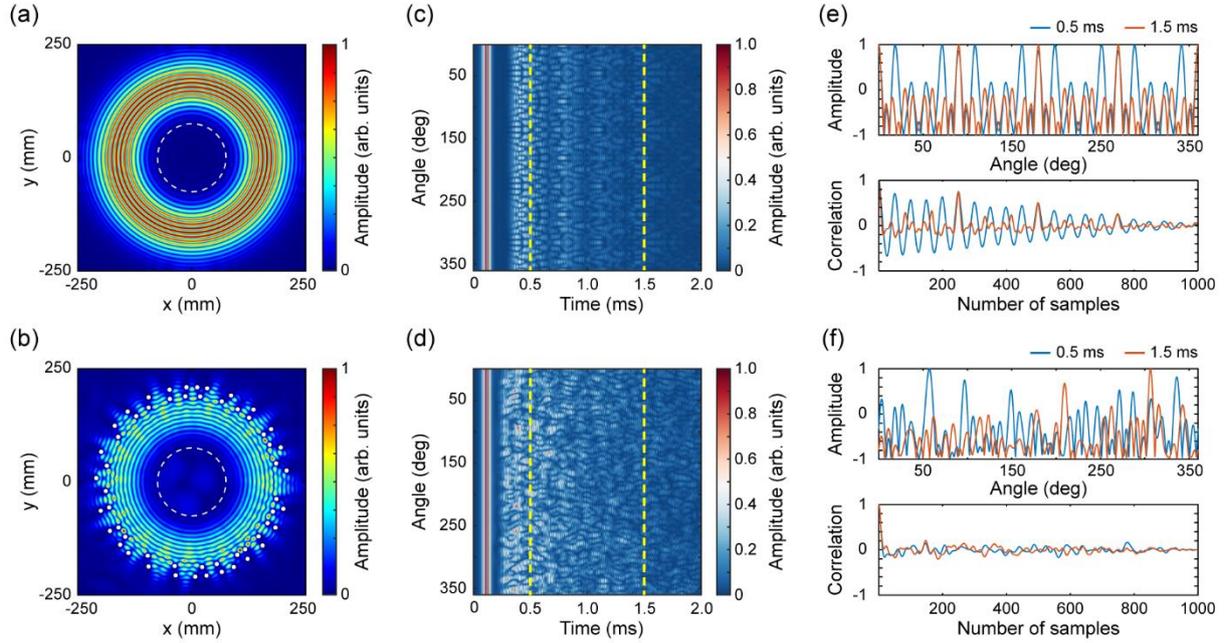

Figure 3. The meta-boundary can break the symmetricity of the time-space evolution property to enable computational identification. The images (a) and (b) show the displacement field distributions at 0.18 msec before and after adding the meta-boundary. The meta-boundary is marked by the white circles. The images (c) and (d) illustrate the time-space evolution diagrams of the displacement fields at the radius of 75 mm (denoted by the white dash lines in images (a) and (b)). The images (e) and (f) show the spatial displacement fields and their autocorrelation functions at 0.5 msec and 1.5 msec (marked by dashed lines in images (c) and (d)) without and with meta-boundary.

**Principle of the spatial scattering coding.** To analyze the spatial scattering coding property of the meta-boundary, we introduce the multiple scattering theory [38] to compute the scattered fields in an infinite thin plate (see Methods). Figure 2(a) shows that the scatterers of the meta-boundary distribute at certain distances $R_1$ (190 mm) and $R_2$ (210 mm) around the center. The scatterers have the diameter of 10 mm and the thickness of 1.5 mm. The angle $\theta$ between each two scatterers is a random variable that determines the positions of scatterers. The incident wave with the frequency of 40 kHz is generated from a point source at the center of the plate. The incident and scattered fields obtained by using the multiple scattering theory are illustrated in Fig. 2(b). From the multiple scattering theory, we can conclude that the spatial scattering coding property of the meta-boundary are determined by the distribution of scatterers. To analyze the influence of scatterers on the spatial scattering coding property, we calculate the scattered fields of interest with the area of $180 \times 180$ mm$^2$ (marked by red square in Fig. 2(a)) by randomly reducing the number of scatterers. We use a correlation index to evaluate the complexity of the scattered field (see Supplementary Note 2). A smaller value of the correlation index reflects larger complexity. With the increase of the number of scatterers, the complexity of the wave field is enlarged and the variation range of the correlation index shrinks (Fig. 2(c)), leading to a desirable highly uncorrelated scattering coding.

To analyze the temporal propagation property of the elastic waves with meta-boundary, we perform the numerical simulation by using the finite element method (see Methods). A Gaussian modulated pulse with the center frequency of 40 kHz is emitted from a transducer fixed on the center of an aluminum plate. The boundaries of the plate are set as the absorption boundaries to suppress the echo. We first calculate the displacement fields of the homogeneous plate at 0.18 msec (Fig. 3(a)). The propagation of the elastic waves is uniform when there are no scatterers. After adding the meta-boundary, strong and uneven reflection occurs when elastic waves encounter the meta-boundary (Fig. 3(b)). More details can be found in Supplementary Video 1. Since the absorption boundary conditions in the numerical simulation cannot fully absorb the echo (which is very common in real environments), the time-space evolution diagram of the displacement fields at the radius of 75 mm without meta-boundary presents a symmetrical pattern due to the weak reflection (Fig. 3(c)). By introducing the scattering-coded meta-boundary, the pattern exhibits disordered characteristics (Fig. 3(d)). To evaluate the characteristics of the patterns, we respectively calculate the autocorrelation functions of the spatial displacement fields at 0.5 msec and 1.5 msec. It can be seen that the autocorrelation functions exhibit periodicity with large values without spatial scattering coding, which reflects the spatial symmetry of the displacement field (Fig. 3(e)). Whereas after introducing the meta-boundary, the autocorrelation functions are close to zero and show randomness (Fig. 3(f)). The results verify that the meta-boundary can still achieve highly uncorrelated scattering coding of elastic waves as time evolves, which is beneficial to identify the temporal and spatial information of objects.

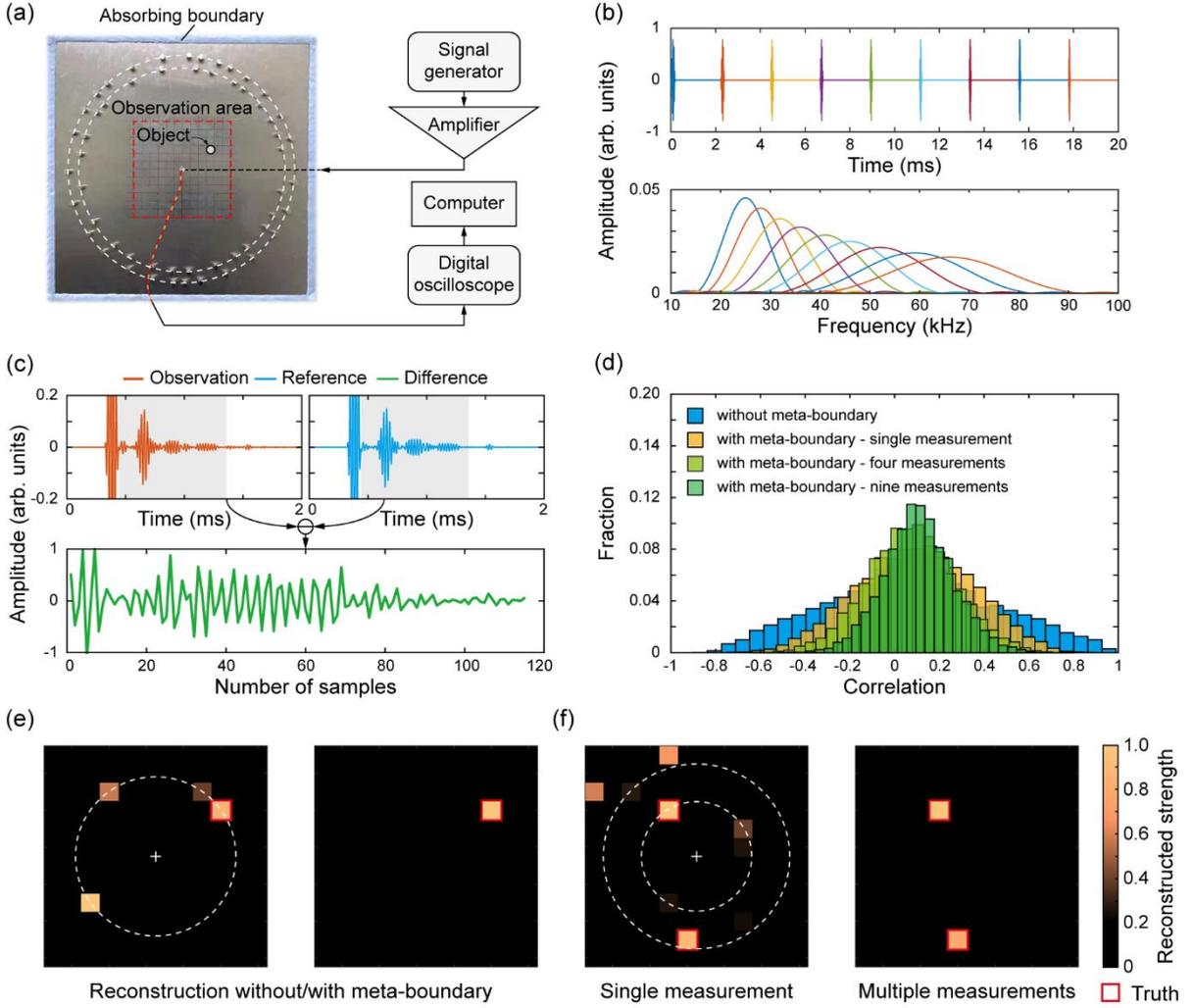

Fig. 4. Performance of the computational identification with meta-boundary. (a) Experimental setup for identifying objects with active elastic waves. (b) Excitation signals with nine different center frequencies. The spectra of the pulses can cover the range from 20 kHz to 75 kHz. (c) The observation signal (with object), reference signal (without object), and their difference signal obtained by a single measurement. (d) The histogram of the cross-correlation coefficients of the difference signals obtained by the calibration process in four cases. (e) Reconstruction results for a single object without and with meta-boundary. The object location can be uniquely identified by introducing meta-boundary. (f) Reconstruction results for two objects with single measurement and multiple measurements.

**Object identification with a single transducer.** To investigate the performance of the meta-boundary for object identification, we conduct experiments on a fixed homogeneous aluminum plate with absorbing boundaries (see Methods).

The meta-boundary is formed by randomly distributed magnets attached at certain distances around the center (Fig. 4(a)). A single piezoelectric transducer is used to generate pulses with different center frequencies (Fig. 4(b)). The frequency range can cover 20 kHz to 75 kHz. A calibration process is performed in advance to obtain the difference signals (Fig. 4(c)) that include the location information of objects (see Methods). The difference signals also reflect the scattering coding property of the meta-boundary. To evaluate the coding performance, we compute the cross-correlation coefficients of the difference signals and illustrate their histograms (Fig. 4(d)). It can be seen that most difference signals are highly correlated without scattering coding. By adding the meta-boundary, the distribution of the cross-correlation coefficients is concentrated towards zero and the high correlations are removed, indicating that the ambiguities of scattered waves are eliminated. The results also illustrate that multiple measurements cause the correlations to be distributed more narrowly around zero, thereby increasing the resolvability of the meta-boundary and improving the performance of the identification. In this way, we demonstrate that the difference signals with multiple measurements can be used to construct a proper measurement matrix for computational identification of objects.

Next, we conduct the testing process by randomly placing one object in the observation area to obtain the observation vector **y**. After that, we use the two-step iterative shrinkage/thresholding algorithm to reconstruct the object vector **x** [39], and reshape **x** into a matrix to obtain a reconstructed image of the observation area. Here, we define it as a correct identification when the highest reconstructed strength matches the actual locations of the objects. Without meta-boundary, the object location is incorrectly reconstructed at the symmetrical points of the actual location (Fig. 4(e)). In contrast, we can uniquely determine the object location with the help of the scattering coding property of the meta-boundary. More comparison results without and with meta-boundary can be found in Supplementary Figs. 2 and 3. The meta-boundary can eliminate the ambiguous information contained in the elastic waves, which is of great significance for localizing multiple objects.

Then, we perform experiments to verify the identification performance for two objects, and compare the reconstruction results obtained by single measurement and multiple measurements (Fig. 4(f)). With single measurement, the first two maximum values in the reconstructed image correctly shows the locations of two objects. However, there are too many noises at other locations, which may lead to the misjudgment of the identification results. By conducting multiple measurements (e.g., nine measurements), the noises at other locations can be well suppressed. Only the reconstructed values of the actual locations can be clearly displayed in the image. More comparison results between multiple measurements and single measurement can be found in Supplementary Fig. 4. The results verify that the meta-boundary is capable of coding spatial elastic waves for computational identification with only a single transducer, which has potential application prospects in various wave sensing fields.

To demonstrate the application prospects of our approach, we design a type of touchscreen based on the computational identification method with meta-boundary. The touchscreen is mainly composed of a tempered glass with meta-boundary and a single piezoelectric transducer as shown in Fig. 5(a) (see Methods for more details). Figure 5(b) shows two samples of the meta-boundary touchscreens, which are used as the devices for entering numbers and letters respectively. We also make two samples of touchscreens without meta-boundary as control groups. We investigate the performance of these touchscreens by touching all characters being pasted on the glass (Figs. 5(c) and 5(d)). It can be seen that the ambiguous information caused by the reflections of the original symmetry boundaries is removed due to the scattering coding of the meta-boundary. The reconstruction results of the meta-boundary touchscreens agree well with the truth, which is crucial for being an input device. Next, we verify the performance of the touchscreens by entering strings "3.1415926" and "shanghai jiao tong university", and visualizing the reconstruction results of the underlined characters (Figs. 5(e) and 5(f)). For control groups without spatial scattering coding, the reconstruction results of some characters are incorrectly reconstructed at the symmetry location of the actual characters, which is intolerable for information input. In contrast, we can accurately identify the strings by spatially coding touchscreen through meta-boundary (see Supplementary Figs. 5, 6, and 7, Supplementary Videos 2 and 3 for more details). The above results demonstrate that the proposed computational identification system with meta-boundary can be used as a new type of human-machine interface for instruction, communication, and encryption without complex hardware and high-power consumption. It can also be flexibly extended to other surfaces and customized according to the practical needs.

**Discussion**

Our proposed design provides a simple but effective strategy for highly uncorrelated scattering coding of elastic waves, which can reduce the hardware complexity of traditional localization methods that rely on transducer arrays. The randomized distribution of the scatterers ensures the uniqueness of the spatial scattering coding, which eliminates the ambiguous information carried by the uncoded scattered waves. Once the distribution of the scatterers is determined, the spatial scattering coding properties are fixed, which means that we only need to perform experimental calibrations once before computational identification. Compared with the existing works of the ultrasonic wave touchscreen [8, 40, 41], our strategy based on the scattering-coded meta-boundary reduces the number of transducers to only one, thereby further reducing the size and power consumption of the device. In addition, the structural configuration and material of the meta-boundary are not strictly limited. We have also demonstrated that the meta-boundary can be migrated to different mediums (e.g., the aluminum plate and the tempered glass). We believe that this universality allows our approach to be applied to a wider range of fields.

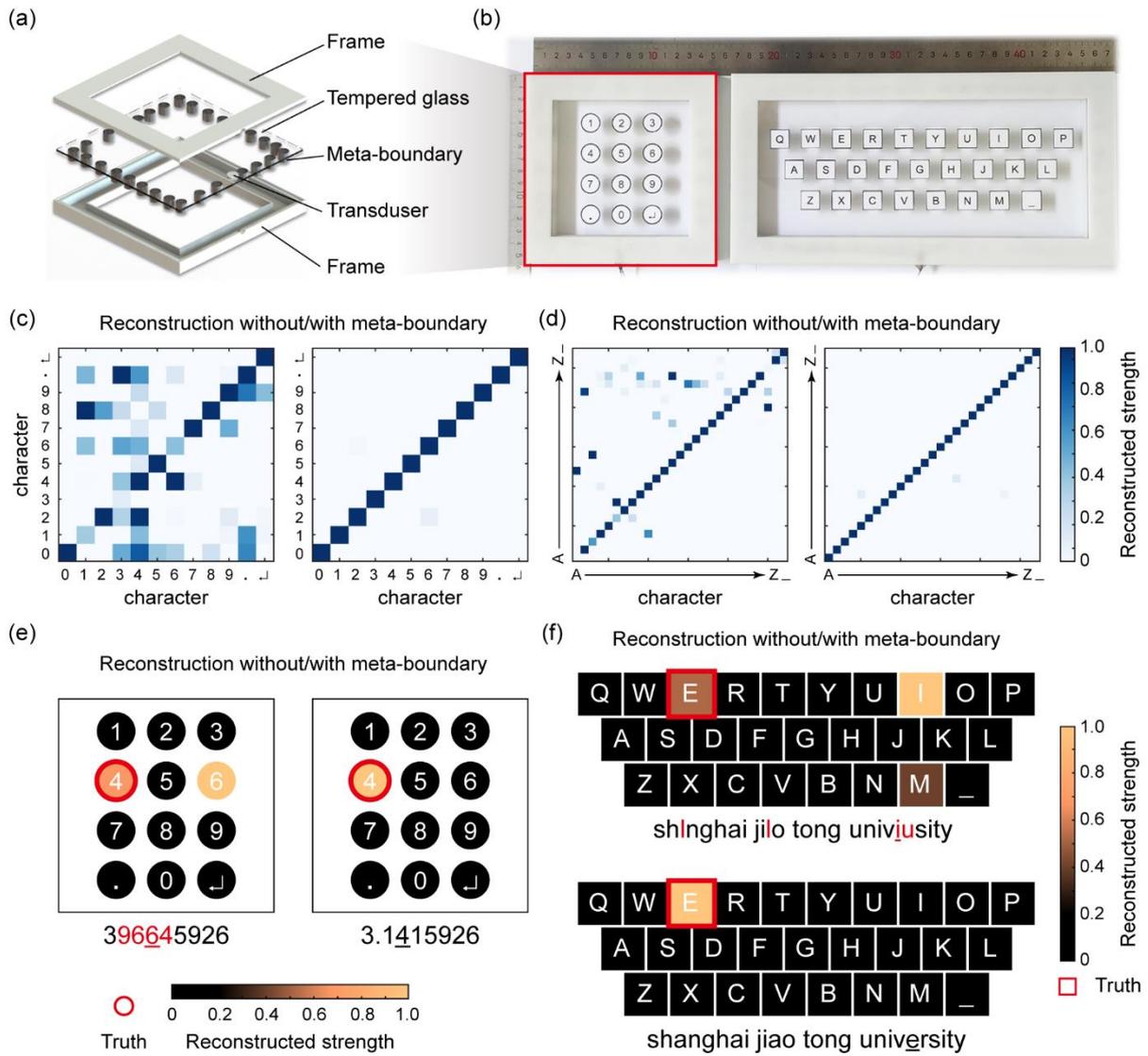

Figure 5. Demonstration of a type of meta-boundary touchscreen. (a) Exploded diagram of the meta-boundary touchscreen. (b) Two samples of meta-boundary touchscreen for entering numbers and letters. Comparison results between the control group and the meta-boundary group for (c) the number keyboard, and (d) the letter keyboard. Reconstruction results for entering (e) numbers "3.1415926" and (f) string "shanghai jiao tong university". The incorrectly identified characters are marked in red color. The demonstration process can be found in Supplementary Video 2 and Video 3.

Although the proposed meta-boundary system is a prototype, it provides an attractive approach for simpler computational identification of elastic waves. There is still room for improvement in the following aspects. The form of the meta-boundary can be designed into holes, gratings, labyrinths, and other types that have strong scattering effects on elastic waves according to actual needs. The theoretical model can be improved to predict the spatial scattering coding of the elastic waves more accurately, so as to reduce the burden of experimental calibration. With the help of advanced manufacturing technology, the integration degree of the meta-boundary system can be improved, thereby reducing the size of the device. By further combing the real-time data acquisition systems and the machine learning algorithms, the meta-boundary devices can realize the trajectory recognition and gesture recognition for more complex human-machine interaction.

In conclusion, we have demonstrated a scattering-coded meta-boundary for computational identification of elastic waves with a single transducer. This work opens up

7 / 10

avenues for artificially designed boundaries with the capability of information coding and identification. We envision that the proposed meta-boundary can be flexibly integrated with various smart devices and platforms (e.g., household appliances, terminal machines, and industrial equipment) to realize an information interaction interface. In addition, we believe that the proposed strategy can provide new perspectives for related wave sensing scenes, such as structural monitoring, underwater detection, indoor localization, and biomedical imaging.

**Methods**

**The multiple scattering analysis of the meta-boundary.** In the multiple scattering analysis, we assume that the scatterers are rigid and have the same thickness as the plate for convenience. For the multiple scattering process, the total scattered field can be regarded as the superposition of the scattered field of each scatterer. In addition to the directly incident wave generated from the source, the incident waves acting on a single scatterer include the scattered waves of other scatterers. Therefore, the scattering coefficient for a scatterer $s$ can be described as

$$\begin{Bmatrix} \mathbf{B}_s^p \\ \mathbf{B}_s^e \end{Bmatrix} = [\mathbf{T}_s] \left( \begin{Bmatrix} \mathbf{A}_s^p \\ \mathbf{A}_s^e \end{Bmatrix} + \sum_{t=1, t \neq s}^{N} \begin{bmatrix} [\mathbf{R}_{ts}^p] & 0 \\ 0 & [\mathbf{R}_{ts}^e] \end{bmatrix} \begin{Bmatrix} \mathbf{B}_t^p \\ \mathbf{B}_t^e \end{Bmatrix} \right), \quad (1)$$

where $[\mathbf{T}_s]$ is the transfer matrix relating the incidence and the scattering of the scatterer, $\{\mathbf{A}_s^p\}$ and $\{\mathbf{A}_s^e\}$ are the coefficient vectors of the directly incident wave, $\{\mathbf{B}_t^p\}$ and $\{\mathbf{B}_t^e\}$ are the coefficient vectors of the scattered waves in the local coordinate system of scatterer $t$. $[\mathbf{R}_{st}^p]$ and $[\mathbf{R}_{st}^e]$ are coordinate transformation matrices between scatterer $s$ and scatterer $t$, given by

$$\left[\mathbf{R}_{st}^p\right]_{mn} = e^{i(m-n)\theta_{st}} H_{m-n}(kd_{st}), \quad (2)$$

$$\left[\mathbf{R}_{st}^e\right]_{mn} = (-1)^n e^{i(m-n)\theta_{st}} H_{m-n}(kd_{st}), \quad (3)$$

where $d_{st}$ and $\theta_{st}$ are the distance and included angle between the two scatterers in the local coordinate system of scatterer $s$, $H_n(\cdot)$ is the Hankel function of order $n$, $k$ is the wave number of the plate. It can be seen that the coordinate transformation matrices are mainly determined by the spatial distribution of scatterers. Once the coefficients of the directly incident waves from the source and the distribution of scatterers are given, the scattered field can be obtained by

$$W^{scr}(r,\theta) = \sum_{s=1}^{N} \left( \{\mathbf{B}_s^p\}^T \{\mathbf{H}(r_s, \theta_s)\} + \{\mathbf{B}_s^e\}^T \{\mathbf{K}(r_s, \theta_s)\} \right), \quad (4)$$

where $(r_s, \theta_s)$ is the representation of the coordinates $(r, \theta)$ in the local coordinate of scatterer $s$, $N$ is the number of scatterers, $\{\mathbf{H}(\cdot)\}$ and $\{\mathbf{K}(\cdot)\}$ are the vector forms of the Hankel function and the modified Bessel function of the second kind. The details of the derivation can be found in Supplementary Note 1. We also demonstrate that the wave scattering effect of the rigid scatterers is highly similar to that of the raised elastic scatterers (see Supplementary Fig. 1(d)), which further illustrates the effectiveness of the multiple scattering theory for analyzing the spatial scattering coding characteristics of the meta-boundary.

**Numerical simulation.** Numerical simulations are conducted to analyze the temporal propagation property of elastic waves by using the piezoelectric interface in the structural mechanics module of COMSOL Multiphysics. The Gaussian modulated pulse with five cycles can be expressed as

$$A_0 \left[1 - \cos(2\pi f_0 t / 5)\right] \sin(2\pi f_0 t), \quad (5)$$

where $A_0$ is the amplitude, and $f_0$ is the center frequency of the pulse. The thickness of the plate is 1.5 mm. The diameter and the height of scatterers are both 10 mm. The distribution of scatterers is the same as that in the multiple scattering analysis. The piezoelectric transducer with the size of $\Phi 10 \times 0.7$ is set as PZT-5H. The material parameters of the plate and the elastic scatterers are set as Young's modulus $E = 70 \times 10^9$ Pa, mass density $\rho = 2700$ kg m$^{-3}$, and Poisson's ratio $v = 0.33$.

**Experiments.** The experimental setup to investigate the performance of the meta-boundary is shown in Fig. 4(a). The dimension of the aluminum plate is set as $500 \times 500 \times 1.5$ mm$^3$. An adhesive called Blu-Tack adheres to the boundaries of the aluminum plate to absorb elastic waves. The piezoelectric transducer is fixed on the center of the aluminum plate by using epoxy adhesive (3M™ Scotch-Weld™ Epoxy Adhesive DP 460). The dimensions and distribution of magnets are the same as that in numerical simulation. Gaussian modulated pulse signals with different frequencies are sequentially generated from a signal generator (SIGLENT Technologies, SDG2122X) and amplified by an amplifier (Nanjing Funeng Technology Industry, HFVA-61) to drive the transducer.

Because the computational identification needs a priori knowledge of the scattering coding property, a calibration process is experimentally performed in advance to construct the measurement matrix. We first pick up the reference signals without placing the object. Then, we successively place a pair of magnets with a diameter of 15 mm and a height of 30 mm as object to be localized on each grid of the observation area in Fig. 4(a), and measure the observation signals. The difference signals are calculated by subtracting the observation signals from the reference signals that contain multiple scattering. The multiple measurements are performed by emitting nine pulses with center frequencies of 25 kHz, 28 kHz, 32 kHz, 36 kHz, 41 kHz, 46 kHz, 52 kHz, 59 kHz, and 66 kHz. The sizes of scatterers and objects are approximately 1/4.6 $\lambda$ and 1/3.1 $\lambda$ at 25 kHz, respectively.

For the touchscreens illustrated in Figs. 5(a) and 5(b), the scatterers with the size of $\Phi 10 \times 8$ are made of 304 stainless steels. The dimensions of two types of tempered glasses are $150 \times 150 \times 2$ mm$^3$ and $300 \times 150 \times 2$ mm$^3$, respectively. The frames are fabricated with photosensitive resin by using light-curing 3D printing.

**Acknowledgements**

This work was supported by the National Natural Science Foundation of China under Grants No. 11872244, No. 52105112, the National Program for Support of Top-Notch Young Professionals, and the China Postdoctoral Science Foundation under Grant No. 2020M680056.

**Author contributions**

Q.H. conceived the idea, proposed the concept of scattering-coded meta-boundary, co-wrote the manuscript, and supervised the entire project. T.J. co-proposed the concept, designed the model of the meta-boundary, performed the numerical simulation and theoretical analysis, carried out the experiments and data analysis, and co-wrote the manuscript. X.L. co-proposed the concept, assisted in the model design, theoretical analysis, experimentation, and data analysis. H.H. performed the theoretical study of the multiple scattering theory and assisted in the experimentation. Z.K.P. supervised the theoretical study and co-wrote the manuscript. All authors discussed the results, commented on, and revised the manuscript.